# Controllable deposition of titanium dioxides onto carbon nanotubes in aqueous solutions

Rongkai Li[a], Dongping Chen[a], Xinling Hu[a], Yanzhen Huang[a], Jing Lu[b], Dongxu Li[a],*

[a] College of Materials Science and Engineering, Huaqiao University, Xiamen 361021, China.

[b] Institute of Manufacture Engineering, Huaqiao University, Xiamen 361021, China.

*Corresponding author. Email: lidongxu@hqu.edu.cn

**ABSTRACT**

Within the field of nanotechnology, nano-scale composites have significant potential in the development of advanced materials for functional applications. Here, composites based on carbon nanotubes and titanium dioxides have been prepared with titanyl sulfate using a chemical bath deposition method at or near room temperature. Two kinds of titanium dioxide depositions corresponding to rutile and anatase were sheathed on carbon nanotubes evenly by adjusting the precursor concentration and temperature. Possible composite mechanisms are discussed. Compared with original tubes, the specific surface areas have been improved nearly four times and the content ratio of mesostructures has also been increased after deposition processes.

# 1. Introduction

Carbon nanotubes (CNTs), as one of the leading one-dimensional nanomaterials, have really attracted wide attention in the past two decades due to their excellent properties and great potential applications [1-3]. These aspects can be further improved or expanded by modifications with designated materials [4-6]. Among them, CNT/metal oxide-based composites are of considerable interest owing to improved combination technology (e.g. self-assembly [7], atomic layer deposition [8]) as well as promising development of functional nano-carbon materials, such as in photocatalysis [9,10], lithium batteries [11,12] and field emission areas [13].

Titanium dioxide ($TiO_2$, titania), beginning with the initial discovery of photocatalytic production of hydrogen under ultraviolet light [14], has been investigated widely at the nanoscale over the past few decades. Generally speaking, there are two main polymorphs of $TiO_2$ in nature, namely, rutile and anatase, which can be ascribed to the difference in the distortion degree of $TiO_6^{2-}$ octahedron [15]. The difference in structures in combination with the diversity of nano-$TiO_2$ morphologies have made its universality in research from photocatalysis, optoelectronic areas [15-18] to biological fields [19]. Although pure $TiO_2$ is endowed with superior physicochemical properties, its limitations in application can be clearly noticed and the construction of composites should extend the scope of titania for aforementioned fields [20]. And considering the characteristics of CNTs (work function (~5 eV) [21], quasi-one-dimentional nanostructure, etc.), CNT/$TiO_2$ composites have become one of the promising hybrids in the field of nanoscience and nanotechnology.

A large amount of efforts have been dedicated to building CNT/TiO$_2$ hybrids through various methods mainly including chemical vapor deposition [22,23] and chemical-solution routes [10,24-28] (e.g. impregnation method, sol-gel and hydrothermal process, self-assembly, and so on). However, most of the exsisting means either require complicated apparatuses or severe reaction environments, and are inclined to the preparation of CNT/anatase composites at selected conditions. Actually, in different applications, the needed surface state and crystal form are not rigid.

In the present work, we show a facile and controllable hydrolyzing process [29,30], which could be classified to chemical bath deposition (CBD) [31], to achieve two forms of TiO$_2$ loaded on CNTs evenly by adjusting the precursor concentrations at relative low temperatures. Titanium oxysulfate–sulfuric acid hydrate (TiOSO$_4$·xH$_2$SO$_4$·xH$_2$O) which is more stable than classical titanium alkoxides or halides at ambient [32], was chosen as Ti precursor. To our knowledge, there is no such systematic research on the different kinds of CNT/TiO$_2$ composites prepared under such mild conditions. The resultant binary composites were characterized by X-ray diffraction, scanning electron microscope, transmission electron microscope, thermogravimetric analysis and nitrogen sorption, *etc*. According to the experimental results, the underlying mechanisms for the formation of different CNT/TiO$_2$ hybrids were proposed.

## 2. Experimental section

*2.1. Preparation of CNT/TiO$_2$ composites.*

The experiment started with the dissolution of 4 g of TiOSO$_4$·xH$_2$SO$_4$·xH$_2$O (Synthesis grade, Aladdin), in 100 mL of distilled water by magnetic stirring (about 11 h at 30 °C), then the solution was filtered to remove insoluble substances. Gradient variations in concentration of the pellucid solution were following established, that is, from 0.04 g/mL to 0.01 g/mL (with pH 0.70→1.20).

In a typical composite experiment, 50 mg of MWCNTs without any pretreatment (diameter 60~100 nm, length>5 μm, from Shenzhen Nanotech Port Co., Ltd.) were added into flasks which contained 50 mL of prepared acidic TiOSO$_4$ solutions above, respectively. After ultrasonically dispersing for about 20 min, the flasks were placed in a thermostatic water bath of 30 °C, and accepted an additional ultrasonic treatment for 5 min every 24 hours. The materials could be filtered and rinsed with amounts of H$_2$O after appropriate reaction time (2~4 days), then were dried at 70 °C overnight under vacuum. To optimize the growth conditions, the deposition temperature was varied in the range of 30 to 60 °C.

*2.2. Sample characterization.*

X-ray powder diffraction (XRD, Smartlab) with Cu $K_α$ radiation was used to identify the phase composition of final products. Surface morphology and size of the composites were observed by field-emission scanning electron microscope (FESEM) (HITACHI S-4800 at 10 kV), transmission electron microscope (TEM) and high-resolution TEM (HRTEM) (TECNAI F30 at 300 kV), respectively. The energy-dispersive X-ray spectroscopy (EDS) data attached to TEM were also recorded. Thermogravimetric analysis (TGA) was carried out in air atmosphere using a DTG-60H instrument (Shimadzu Co., Japan) with a heating rate of 10 °C min$^{-1}$. Low-pressure (up to 1 bar) gas adsorption/desorption isotherms ($N_2$) were measured on a Micrometrics ASAP 2020 surface area and porosity analyzer.

## 3. Results and discussion

After the hydrolyzing experiments around the concentrations and temperatures, the phase compositions of the black-gray products were identified by XRD. To visualize the phase transition process of $TiO_2$, we roughly calculated the relative content of the rutile phase in the hybrids from the XRD peak intensities by an experienced formula [33]:

$$\omega_R = 1 / [ 1 + 0.884 (a_A / a_R) ] \qquad (1)$$

where $a_A$ and $a_R$ represent integrated intensities of anatase (101) and rutile (110) diffraction peaks, respectively.

As shown in Fig. 1, pure rutile titania was prepared under the initial experimental condition (0.04 g/mL, 30 ℃), in which only peaks of rutile phase were observed except the graphite peak of CNTs in the XRD curve (line a in Fig. 2). The crystal phase of $TiO_2$ in the hybrids gradually transformed from rutile to anatase when precursor concentration decreased (meanwhile pH increased because of the contained $H_2SO_4$ in the precursor). This trend was also emerged with temperature elevating. Finally, under the condition (0.01 g/mL, 60 °C) as well as its next experimental points in the distribution map, pure anatase phase appeared on the stage. The line b in Fig. 2 shows the diffraction peaks of the generated anatase $TiO_2$ along with CNTs. The displayed peaks of CNTs are blur, which can be possibly attributed to that the main characteristic peak of CNTs at 26.1° is overlapped by the (101) peak at 25.2° of anatase $TiO_2$ [9].

In this work, we mainly focused on the pure crystal phase of $TiO_2$ combined with CNTs. The products from the reaction condition of (0.04 g/mL, 30 °C) and (0.01 g/mL, 60 °C) would be chosen as representative samples for CNT/rutile and CNT/anatase composites, respectively, and named R and A as a matter of convenience.

The prepared CNT/rutile and CNT/anatase composites can be observed in FESEM images, respectively in Fig. 3(b) and (d). Fig. 3(b) indicates the presence of rutile $TiO_2$ attaching to the CNTs homogeneously in a hump-like shape, when comparing with the original CNTs in Fig. 3(a). As for deposited anatase $TiO_2$ (Fig. 3d), conformal coatings wrapping CNTs are the main morphology. Fig. 3(c) shows a representative microscopic morphology of CNT/mixed rutile-anatase phase composites, of which reproduces the both constructions of the two forms of titania on CNTs.

To further uncover the microtopographies and nanostructures of the two kinds of CNT /$TiO_2$ composites, their TEM and HRTEM images are investigated (Fig. 4). The loading rutile can be described as cluster-like deposition as shown in Fig. 4(a) and (b). The orientation of the so-called clusters is random, and the polycrystallinity of the material matches well with the selected area electron diffraction (SAED) pattern (inset in Fig. 4(a)). As the tubular material shown in Fig. 4(d) and (e), the continuous coating (25~30 nm) sheathed on CNTs can be clearly seen. The SAED picture (inset in Fig. 4(d)) of the TiO$_2$ layer indicates that the structure corresponds to anatase. Based on HRTEM characterization of CNT/TiO$_2$ composites in Fig. 4(c) and (f), the (110) and (101) lattice plane corresponding to rutile and anatase, respectively, can be apparently observed in spite of their scattered distributions. Additionally, the Ti/O atomic ratios (about 1:2) in both composites were verified by EDS analysis. In principle, the information from TEM is consistent with XRD spectrums.

According to the above characterizations of samples from controlled conditional experiments, possible mechanisms are proposed to elucidate the formation of the two kinds of CNT/TiO$_2$ hybrids (Fig. 5). When CNTs dispersing in the Ti precursor solution, the tube surfaces are encompassed by soluble components. Precursor substances would absorb on the exterior walls of tubular templates due to the large specific surface area of CNTs. With appropriate time of reaction, precursors continuously condense to TiO$_2$ crystal nucleuses and grow at a given temperature. With regard to precise crystal configurations of titania, the partial charge model is generally cited to explain the linkage of TiO$_6^{2-}$ octahedron [34]. Typically, the starting Ti precursor would be hydrolyzed to [Ti(OH)(OH$_2$)$_5$]$^{3+}$ complex at first, then formed to [Ti(OH)$_2$(OH$_2$)$_4$]$^{2+}$, [TiO(OH)(OH$_2$)$_4$]$^+$, [Ti(OH)$_3$(OH$_2$)$_3$]$^+$ or [Ti(OH)$_4$(OH$_2$)$_2$]$^0$ on the basis of the pH and temperature of the solution [35,36]. Lower acidity or elevated temperature tends to promote the hydroxylating process within a certain range. According to the induction of Girish Kumar et al. [34], we suggest that Ti precursors with low degree of hydroxylation are prone to linear growth along the equatorial plane of $Ti^{4+}$ ions, which can only form rutile nuclei. In contrast, condensation would grow along the apical directions, leading to distorted chains of anatase.

As stated in the introduction, one feature in this study is the use of $TiOSO_4 \cdot xH_2SO_4 \cdot xH_2O$ as Ti precursors in the reaction system. The chemical formula indicates that $H_2SO_4$ is one of the existing components in the raw material. Actually, the included $H_2SO_4$ plays an important role in the whole experiments. On one hand, $TiOSO_4$ is insoluble in the solvents at higher pH values [32]. On the other hand, the formation of rutile or anatase closely relates to acidity more than the concentration of Ti precursor itself [32,34]. So as to achieve transparent precursor solution and featured phase deposition, we regulate the acidity directly by adjusting the amount of our raw materials without additional acid.

In aspect of the morphology of crystallites, the reaction rate is presumed to be a key factor [32]. The extent of reaction is estimated by TGA of the two forms of $CNT/TiO_2$ composites (Fig. 6) except the slight color change rate of the solution.

The residual mass fractions of R and A after removing burnable CNTs are about 78.1% and 67.9%, respectively. However, the initial amounts of $TiOSO_4$ in R is quadrupled A. This result demonstrates the faster reaction rate in production of CNT/anatase than CNT/rutile in this aqueous system. Yamabi et al. [32] deemed that a relatively low reaction rate mainly leaded to heterogeneous nucleation and generates more oriented crystals, cluster-like rutile in this work. On the contrary, higher reaction rate would result in the direct deposition of non-oriented anatase phase via homogeneous nucleation.

Next, $N_2$ adsorption/desorption isotherms were adopted to explore the difference in surface characteristics of the tubes before and after coating CNTs by $TiO_2$ in Fig. 7. The nitrogen isotherms and pore size distribution line (Fig. 7a) of the adopted CNTs exhibit an obvious $N_2$ absorption at pore sizes of ~40 nm which can be ascribed to the filling of the CNTs voids and of spaces in tube agglomerates [37,38]. After deposition with titania coverings, the content of small mesopores (2~5 nm) are a majority, which can be observed from the inset of Fig. 7b and 7c. In accordance with the $N_2$ adsorption/desorption isotherms, both of the composites exhibit much higher BET surface areas (R: 263 $m^2/g$, A: 226 $m^2/g$) than CNTs (69 $m^2/g$). However, only the curve of CNT/anatase composites show the type IV isotherms with a depressed hysteresis loop (Fig. 7c) which is generally corresponding to mesoporous structures [39], and agrees well with its pore size distribution plot. No obvious similar hysteresis loop is found in R, possibly because of the more oriented growth of rutile than the uniformly accumulated anatase particles. To sum up, cluster-like rutile depositions are conductive to improving the surface areas of CNTs, while flat anatase coating in favor of adding mesoporous structures for CNTs.

## 4. Conclusions

In this paper, we have demonstrated a simple, low-temperature, solution-based method to fabricate different forms of CNT/$TiO_2$ composites by adjusting the precursor concentrations and temperatures of the system. Various characterizations were carried out from phase structures to morphologies of the prepared composites, and revealed the respective depositing ways of rutile and anatase onto CNTs. To better understand the influence of the controllable variables on crystal phases and formative morphologies, we also explored the underlying mechanisms. In addition, by depositing $TiO_2$ on CNTs, the specific surface areas and contents of mesopores of the tubes have been improved obviously. It is expected that these $CNT/TiO_2$ composites, whether used directly or through further postprocessing, have great potential for the existing catalytic, optoelectronic or other areas. Meanwhile, this work has enriched the approaches of building CNT/metal oxide composites and $TiO_2$-based hybrids.

**Acknowledgments**

This work was supported by National Natural Science Foundation of China (No.51502098 and 51475175), and Promotion Program for Young and Middle-aged Teacher in Science and Technology Research of Huaqiao University (ZQN-PY30). The authors also appreciate help and advice from Prof. Jincao Dai and Genggeng Luo.

**Figure Captions**

**Fig. 1** Distribution relationship between the prepared conditions (precursor concentration and temperature) and the deposited crystal phases on CNTs. The number subscript near R or A is its relative content in a mixed rutile-anatase phase.

**Fig. 2** XRD spectrums for CNT/TiO$_2$ hybrids prepared at two different conditions: (a) 0.04 g/mL TiOSO$_4$ (pH 0.70) and 30 °C for CNT/rutile, (b) 0.01 g/mL TiOSO$_4$ (pH 1.20) and 60 °C for CNT/anatase.

**Fig. 3** FESEM images. (a) Pristine CNTs, (b) The CNT/rutile composites, (c) Representative samples with mixed rutile-anatase on CNTs, (d) The CNT/anatase composites.

**Fig. 4** (a) TEM image of CNT/rutile composites with the SAED pattern. (b) Partial enlargement of a. (d) TEM image of anatase TiO$_2$ coated CNT (inset: SAED pattern). (e) Partial enlargement of d. (c and f) HRTEM images of CNT/rutile and CNT/anatase hybrids, respectively, with their Fourier transforms (inset).

**Fig. 5** Schematic diagram of the preparation of CNT/TiO$_2$ composites.

**Fig. 6** TGA curves of CNT/TiO$_2$ composite samples (R and A) and original CNTs. The black line shows that CNTs can be burnt out at about 720 °C. The amount of weight loss in composites indicates the percentage content of TiO$_2$ (R: 78.1%; A: 67.9%).

**Fig. 7** Nitrogen adsorption/desorption isotherms of (a) pristine CNTs, (b) R and (c) A. Inset in each picture shows the pore size distribution plot. The BET surface areas are 69 m$^2$/g, 263 m$^2$/g and 226 m$^2$/g,

respectively.

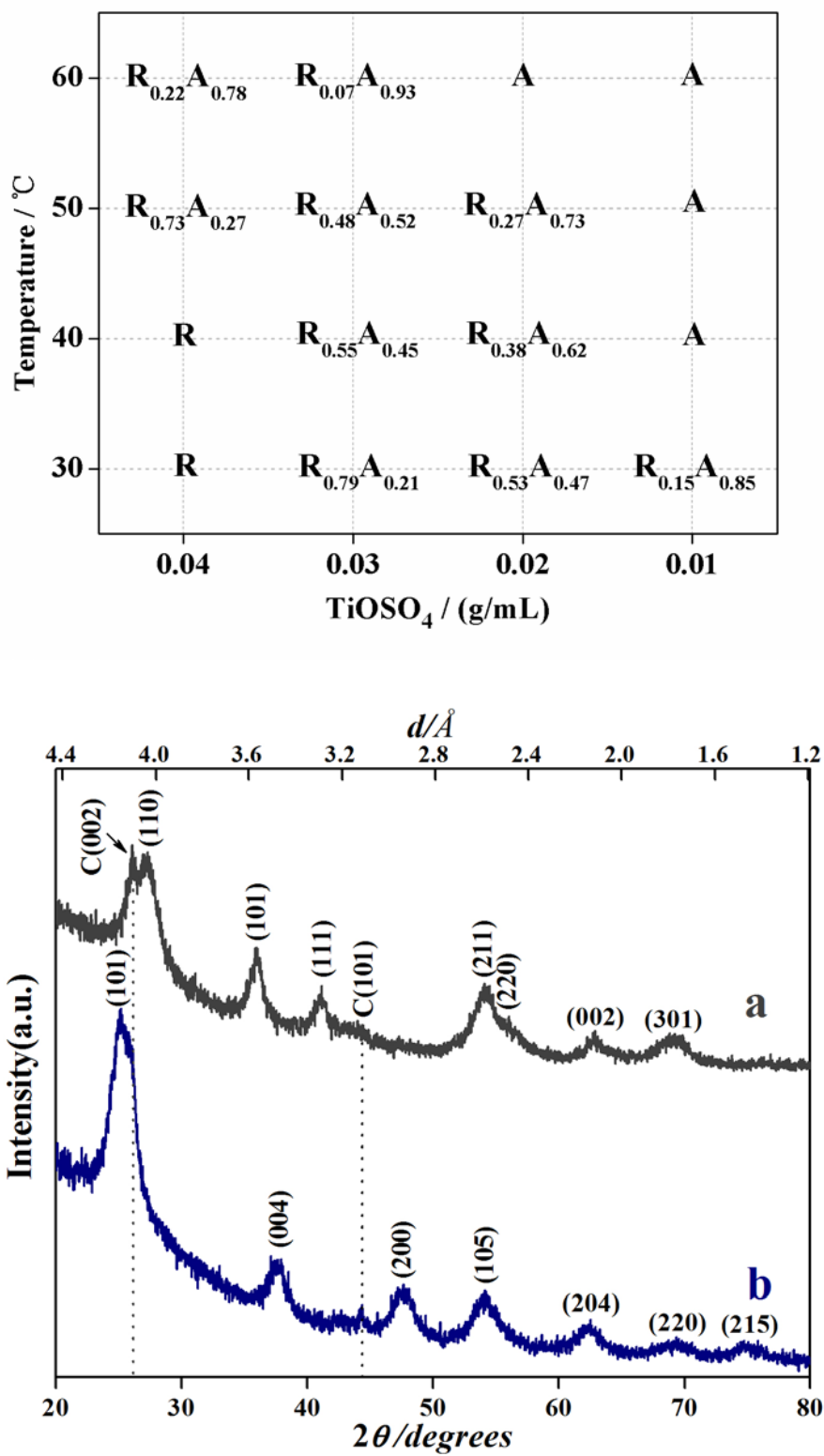

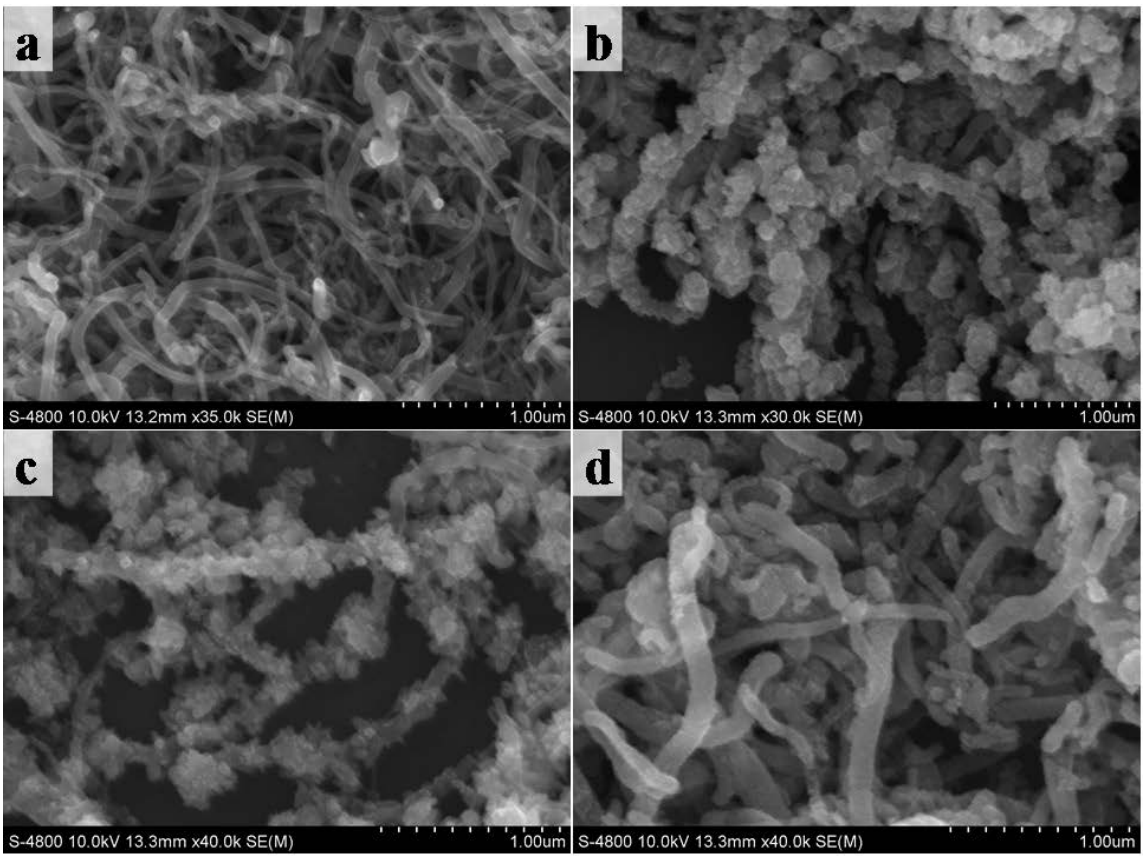
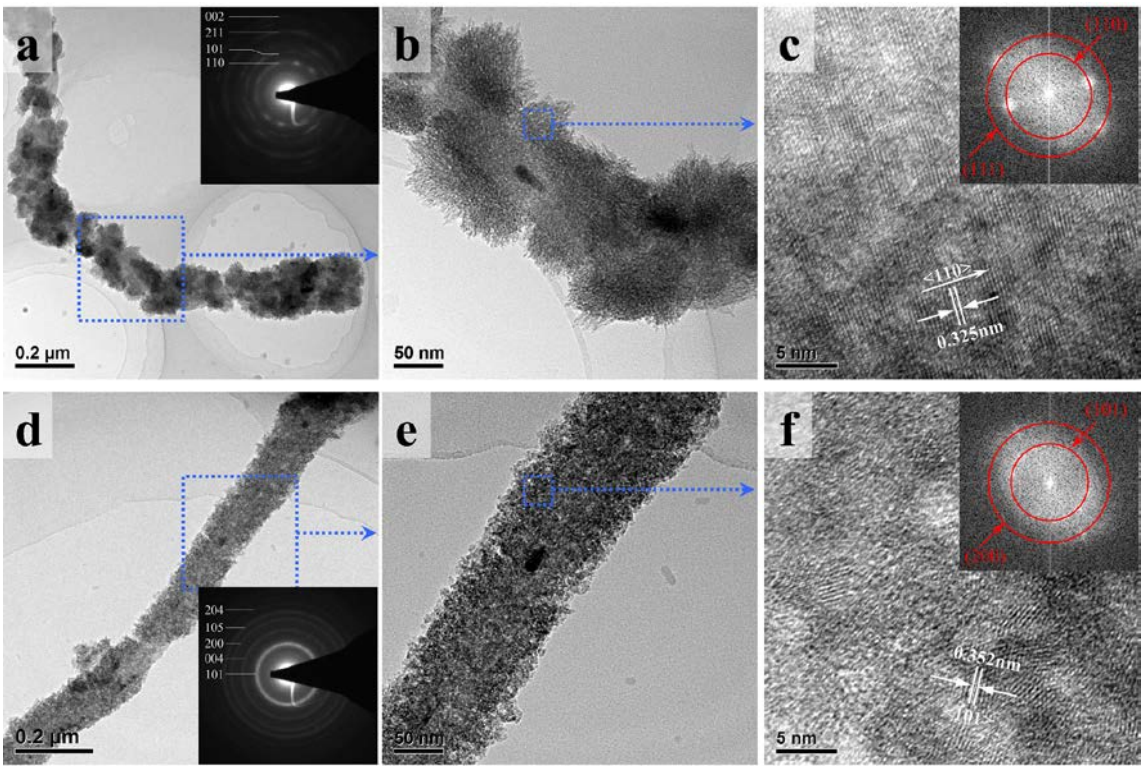

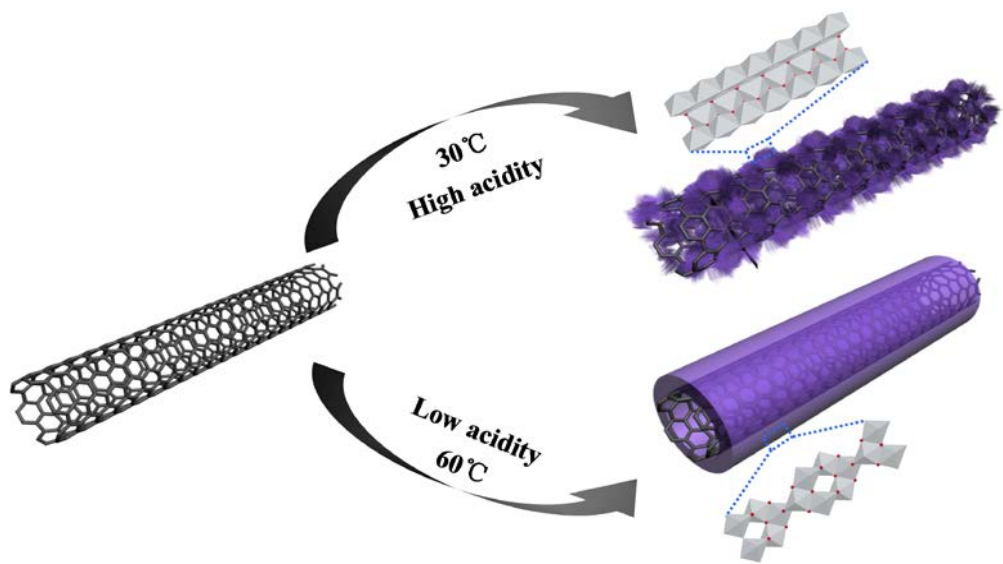

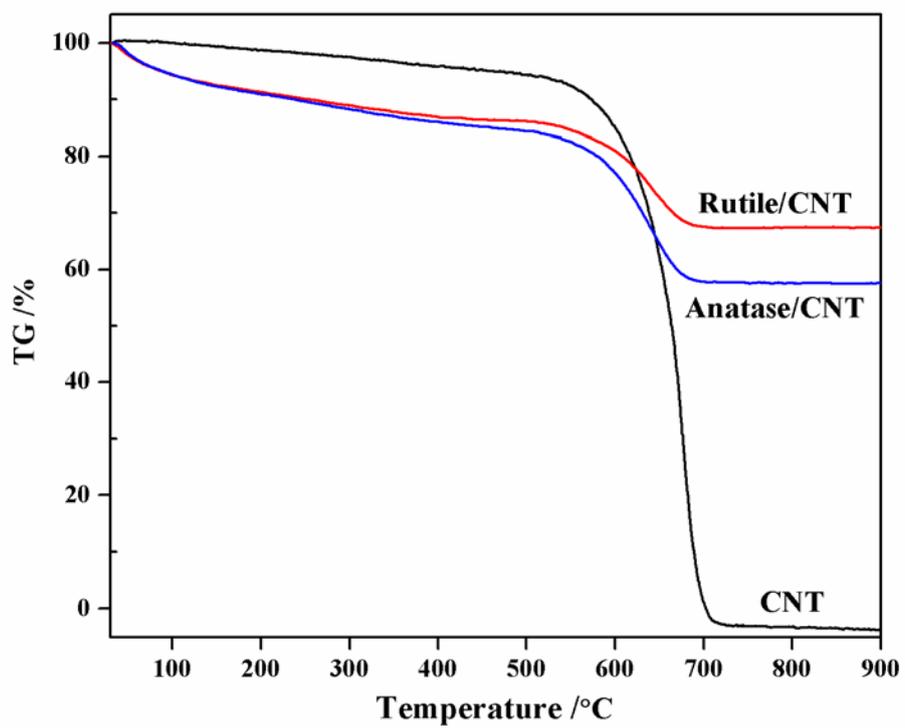

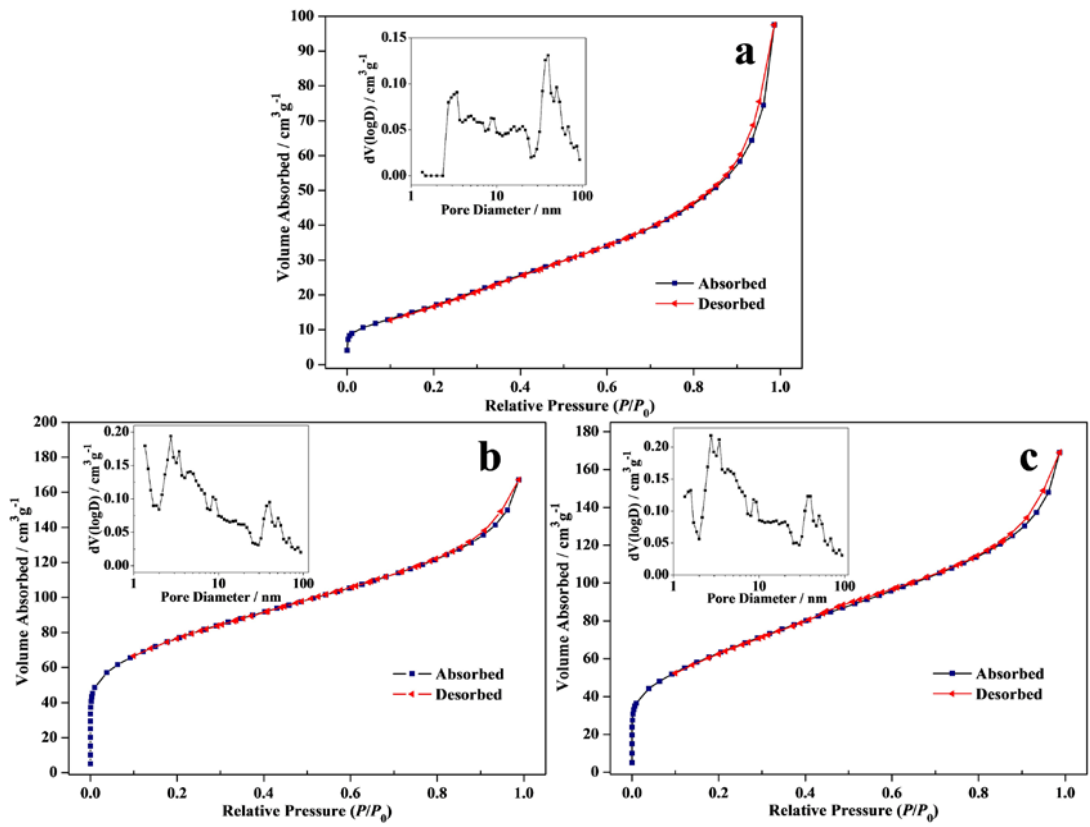